\documentclass[conference,letterpaper]{IEEEtran}

\usepackage[utf8]{inputenc} 
\usepackage[T1]{fontenc}
\usepackage{url}              
\usepackage{cite}             

\usepackage[cmex10]{amsmath}  

\usepackage{mleftright}       
\mleftright                   

\usepackage{graphicx}         
\usepackage{booktabs}         

\usepackage{soul}

\usepackage{xcolor}

\allowdisplaybreaks[2]
\graphicspath{{figures/}}

\usepackage{amsthm}
\usepackage{amsmath}
\usepackage{amsfonts}
\usepackage{mathtools}
\usepackage{cancel}

\newtheorem{theorem}{Theorem}

\usepackage{subcaption}

\usepackage{cleveref}
\crefname{equation}{Eq}{Eqs} 
\crefrangelabelformat{equation}{(#3#1#4--#5#2#6)}

\DeclarePairedDelimiter{\nint}\lfloor\rceil

\usepackage{float}

\usepackage{tikz}
\usetikzlibrary{backgrounds, calc, chains, fit, matrix, positioning, shadows, shapes, intersections, circuits.ee}
\tikzset{input/.style={}}
\tikzset{output/.style={}}
\tikzset{op/.style={circle, draw, fill=black!10, minimum size=2.5ex, inner sep=0ex}}
\tikzset{filter/.style={rectangle, draw, thick, fill=black!10, minimum size=3.5ex, inner sep=1ex}}
\tikzset{nn/.style={trapezium, trapezium angle=80, draw, thick, fill=black!10, inner sep=1ex}}
\tikzset{branch/.style={circle, draw, thick, fill=black, minimum size=.5ex, inner sep=0ex}}
\tikzset{tensor/.style={rectangle, draw, fill=white, minimum size=2em, double copy shadow={shadow xshift=.5ex,shadow yshift=-.5ex}}}
\tikzset{rounded/.style={rounded rectangle, draw, thick, fill=black!10, minimum size=3.5ex, inner xsep=1ex}}
\tikzset{image/.style={rectangle, draw, fill=white, minimum size=2em}}
\tikzset{>=direction ee}
\tikzset{/tikz/thin/.style={line width=.9pt}}
\tikzset{/tikz/thick/.style={line width=1.4pt}}
\tikzset{every path/.style={thin}}

\usepackage{pgfplots}
\pgfplotsset{compat=1.14}
\pgfplotsset{every axis/.append style={enlargelimits={abs=3pt},grid,axis lines=left}}
\pgfplotsset{every axis plot/.append style={thick,mark size=1.5pt,line join=bevel,mark options={solid}}}
\pgfplotsset{label style={font=\small}}
\pgfplotsset{tick label style={font=\footnotesize}}
\pgfplotsset{grid style={color=black!10}}
\pgfplotsset{legend style={draw=none,opacity=.85,font=\footnotesize,cells={anchor=west,opacity=1}}}
\pgfplotsset{every non boxed x axis/.style={xtick align=center,shorten <=-.5\pgflinewidth}}
\pgfplotsset{every non boxed y axis/.style={ytick align=center,shorten <=-.5\pgflinewidth}}
\pgfplotsset{every non boxed z axis/.style={ztick align=center,shorten <=-.5\pgflinewidth}}
\pgfplotsset{/pgf/number format/1000 sep={\,}}

\usepackage[font=small]{caption}

\IEEEoverridecommandlockouts

\begin{document}

\title{Distributed Compression in the Era of Machine Learning: A Review of Recent Advances} 

\author{%
  \IEEEauthorblockN{Anonymous Authors}
  \IEEEauthorblockA{%
    Please do NOT provide authors' names and affiliations\\
    in the paper submitted for review, but keep this placeholder.\\
    ISIT23 follows a \textbf{double-blind reviewing policy}.}
}

%
%
\author{%
  \IEEEauthorblockN{Ezgi~{\"O}zyılkan and Elza Erkip}
  \IEEEauthorblockA{Dept.~of Electrical and Computer Engineering \\
  New York University \\
  NY 11201, USA \\
\texttt{\{ezgi.ozyilkan, elza\}@nyu.edu}}
}

\maketitle

\begin{abstract} Many applications from camera arrays to sensor networks require efficient compression and processing of correlated data, which in general is collected in a distributed fashion. While information-theoretic foundations of distributed compression are well investigated, the impact of theory in practice-oriented applications to this day has been somewhat limited. As the field of data compression is undergoing a transformation with the emergence of learning-based techniques, machine learning is becoming an important tool to reap the long-promised benefits of distributed compression. In this paper, we review the recent contributions in the broad area of learned distributed compression techniques for abstract sources and images. In particular, we discuss approaches that provide interpretable results operating close to information-theoretic bounds. We also highlight unresolved research challenges, aiming to inspire fresh interest and advancements in the field of learned distributed compression.

\end{abstract}


\section{Introduction}
\label{sec:intro}

Data compression, also known as source coding, is a fundamental and extensively studied problem in both information theory and engineering disciplines. Shannon established that the \emph{entropy} of a source serves as a fundamental limit of lossless data compression~\cite{Shannon1}. Achieving lossless compression near the Shannon limit requires that multiple samples of the information source to be compressed jointly, allowing for a small but vanishing probability of error as the blocklength increases. Counterpart to his lossless source coding theorem, Shannon's \emph{lossy} source coding theorem~\cite{Shannon1, Shannon2} provides theoretical limits on how efficiently one can compress an information source while allowing for a given level of \emph{distortion} in the reconstruction. The so-called rate--distortion theory is particularly relevant in practical scenarios where the source is often continuous-valued, such as vectors representing image pixel intensities. For such sources, it is common in practice to initially quantize the data to a finite set of discrete values, thereby introducing a certain level of error. This gives rise to the challenge of lossy compression problems, where a trade-off must be struck between two competing goals: the entropy of the discretized representation (rate) and the error stemming from quantization (distortion).

The optimal trade-off in lossy compression is determined by information-theoretic rate--distortion function~\cite{elements_of_information_theory}, which is fully characterized only for specific source distributions. Given the inherent complexity of lossy compression, practical codes are often tailored to specific domains, such as image, audio, or video. In the context of lossy image compression, well-known codecs typically adopt a two-step process: first, a linear transformation followed by quantization, and then lossless compression using entropy coding, in the form of arithmetic coding, to efficiently transmit the induced quantized representation. This two-step approach is exemplified in widely used standards like JPEG~\cite{JPEG} and JPEG 2000.

In recent years, there has been a significant surge in research on data-driven image compression algorithms enhanced by deep neural networks (DNN)~\cite{Balle2017, Balle2018, BalleJournal, Minnen2018}. These methods have demonstrated impressive performance, surpassing classical standard compression techniques like JPEG 2000 and BPG~\cite{BPG}. As argued in~\cite{Balle2017, BalleJournal}, the effectiveness of DNN-aided compressors lies in their use of learned \emph{nonlinear} transforms, aligned with the universal function approximation capability of artificial neural networks (ANNs)~\cite{Leshno}. Unlike traditional methods that rely on handcrafted representations (e.g., discrete cosine transform as in the JPEG algorithm) to capture natural image statistics, neural compressors learn the underlying representations in an end-to-end fashion directly from the data itself.

Distributed source coding (DSC) addresses the challenge of efficiently compressing information from physically separated sources that are then decoded jointly. High correlation between distributed data streams naturally arises in many current practical setups, such as camera arrays and sensor networks. While the theory of DSC predicts substantial improvements in compression efficiency compared to the point-to-point setup~\cite{elements_of_information_theory}, practical distributed compressors have not been fully developed to this day mainly due to the difficulty of handling complex correlations between information sources. We posit that learning-based methods could be helpful in this regard, given their quick adaptability to arbitrary data sources and to new modalities of multimedia.

The goal of this review article is to summarize recent results in the literature on DNN-aided DSC, with a specific focus on a special and simpler distributed lossy compression case where the side information is available only at the decoder side, also known as the Wyner--Ziv problem~\cite{Wyner_Ziv}. To understand the efficacy of learning-based techniques and the conditions under which they perform competitively, we also discuss setups involving abstract sources in addition to practice-oriented ones that involve images.

The subsequent sections of this paper are structured as follows. Section~\ref{sec:theoretical_background} offers a brief overview of information-theoretic foundations of distributed data compression. Transitioning to neural/learned compression schemes, we recap the current state-of-the-art framework in Section~\ref{sec:NTC}. Section~\ref{sec:dsc_real_life} delves into practical examples of various learned DSC scenarios, mostly involving images as information sources, and identifies gaps in the deep learning literature. Section~\ref{sec:learned_wz} provides a compilation of recent neural distributed compression techniques considering decoder-only side information, with a focus on abstract sources. Lastly, Section~\ref{sec:conclusion} outlines avenues for future research.

\section{Information-Theoretical Background}
\label{sec:theoretical_background}

In this section, we provide the well-known result on lossless compression of separately encoded sources, often referred to as the \textit{Slepian--Wolf} theorem. Subsequently, we shift our focus to distributed lossy compression, specifically considering the scenario with decoder-only side information that is known as the \emph{Wyner--Ziv} theorem.

\subsection{Lossless Distributed Source Coding (Slepian--Wolf)} \label{sec:SW}

The lossless distributed compression problem was first studied by Slepian and Wolf for two sources~\cite{Slepian_Wolf}. The main result is presented below. For achievability and converse parts of the theorem, we refer the readers to~\cite{network_info_theo} as well as the original paper in~\cite{Slepian_Wolf}.

\begin{figure}[H]
    \centering
\includegraphics[width=0.9\linewidth]{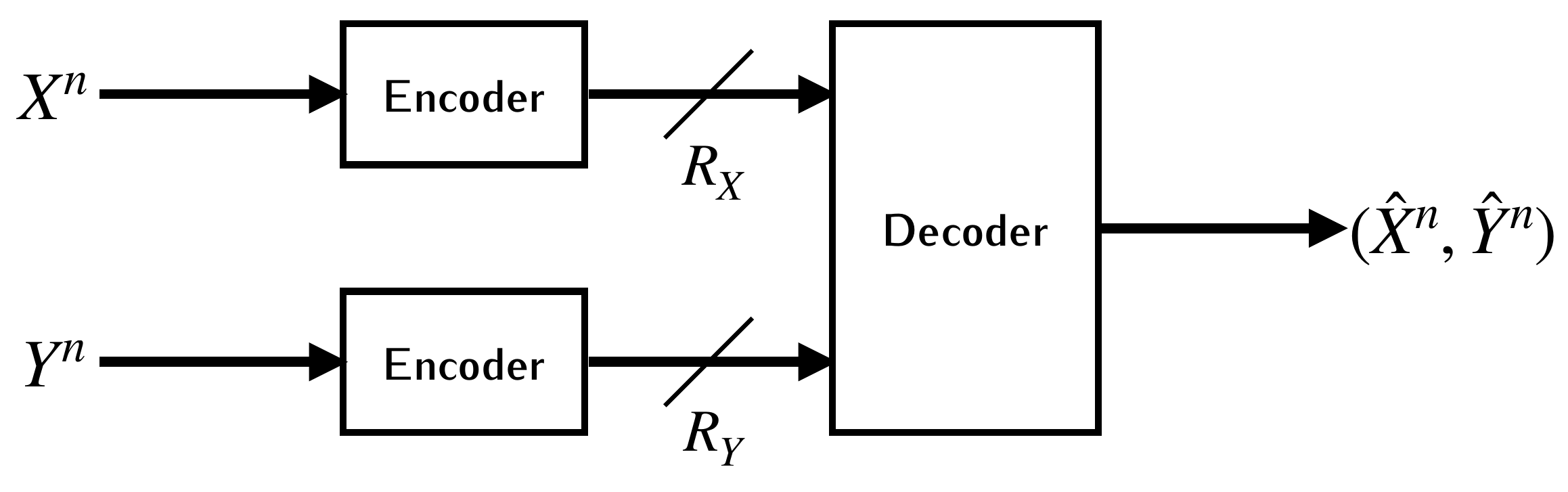}
  \caption{Distributed lossless compression with two sources, also known as \emph{Slepian--Wolf} coding.}
  \label{fig:SW}
\end{figure}

\begin{theorem} \label{theo:1} (Slepian--Wolf Theorem [1973]) The optimal rate region for distributed lossless source coding of a pair of discrete memoryless sources $(X,Y)$ is the set of pairs $(R_{X}, R_{Y})$ such that:
\begin{align}
    R_{X} & \geq H(X \mid Y), \label{eq:SW1} \\
    R_{Y} & \geq H(Y \mid X), \\
     R_{X} + R_{Y} & \geq H(X, \: Y)  . 
\end{align}
\end{theorem}

\noindent The achievability part of the theorem involves a coding procedure using \emph{random bins}, due to Cover~\cite{Cover1}. As in many information theory problems, the achievability assumes asymptotically large blocklengths. The converse part invokes Fano's inequality and chain rule arguments. Note that the Slepian--Wolf theorem can be straightforwardly extended to distributed lossless compression with an arbitrary number of sources~\cite{elements_of_information_theory}.

Let's now consider a corner point of the achievable Slepian--Wolf rate region, with rate $R_{Y} = H(Y)$, $R_{X} = H(X \mid Y)$. This operationally corresponds to the scenario that the source~$Y$ is losslessly available at the decoder side. Surprisingly, one can observe that an encoder that does not have access to the correlated side information can asymptotically achieve the same compression rate with the case when side information is available at both the encoder and the decoder. In other words, two distributed (or isolated) encoders can losslessly compress their sources as efficiently as if they were communicating with one another.

\subsection{Lossy Source Coding with Side Information (Wyner--Ziv)}\label{sec:WZ}

A general distributed lossy compression setup involving two encoders and two distortion measures was studied by Berger and Tung~\cite{multiterminal}. The rate region of the quadratic-Gaussian setup involving two encoders was explicitly characterized in~\cite{Aaron_quadratic}. Here, we only focus on the Wyner--Ziv setup as it captures the simplest distributed lossy compression setting, assuming decoder-only side information. For the achievability and converse parts of the theorem, we refer the readers to~\cite{network_info_theo}, as well as the original paper in~\cite{Wyner_Ziv}.

\begin{figure}[H]
    \centering
  \includegraphics[width=0.9\linewidth]{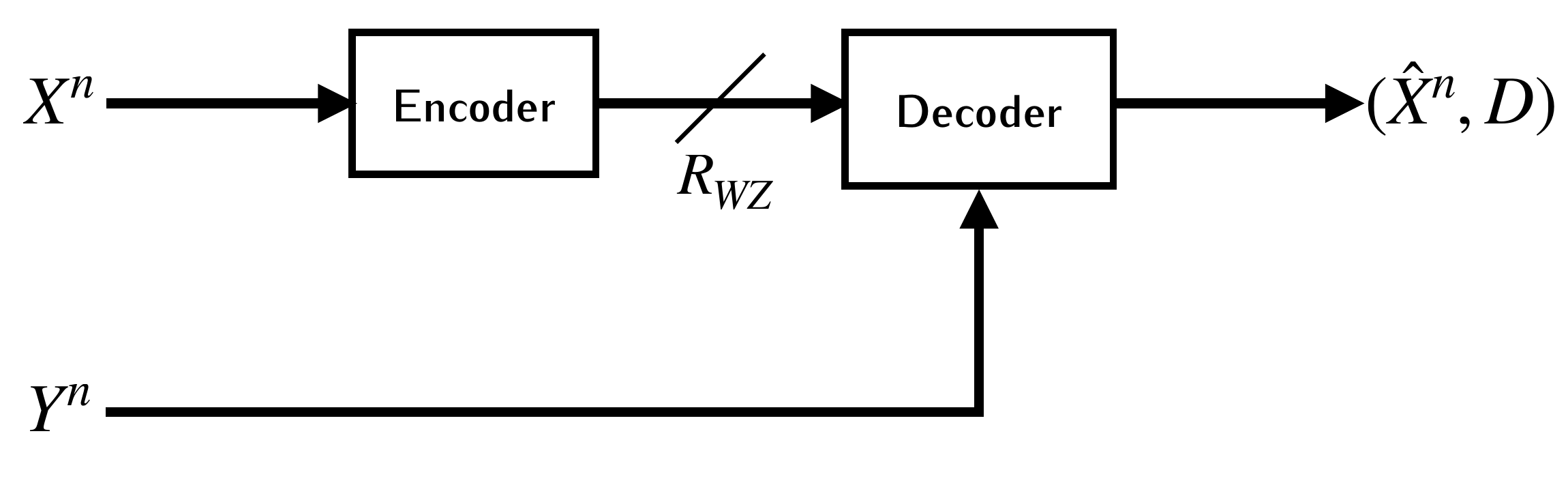}
  \caption{Rate--distortion with side information, also known as \emph{Wyner--Ziv} coding.}
  \label{fig:WZ}
\end{figure}

\begin{theorem} \label{theo:WZ}(Wyner--Ziv Theorem [1976]) Let $(X,Y)$ be correlated sources, drawn i.i.d. $\sim p(x,y)$, and let $d(x, \hat{x})$ be a single-letter distortion measure. The rate--distortion function for $X$ with side information $Y$
available (non-causally) at the decoder side is as follows:
\begin{equation} \label{eq:WZ}
    R_{\text{WZ}}(D) = \min (I(X;U) - I(Y;U)),
\end{equation}
where the minimization operation is over
all conditional probability distribution functions $p(u \vert  x)$,
and all functions $g(u,y)$ such that  $ \mathbb{E}_{p(x,y)p(u\vert x)} d(x, g(u,y) ) \le D$. \end{theorem}The achievability part of the Wyner--Ziv theorem, which again relies on asymptotically large blocklengths, invokes the covering lemma, resulting in the rate of $I(X;U)$, followed by a random binning argument based on joint typicality, which yields the rate discount of $I(Y;U)$ in Eq.~\eqref{eq:WZ}~\cite{elements_of_information_theory}. This achievability, which is shown to be tight, assumes a Markov chain constraint $U-X-Y$. 

In general, the optimal $p(u \vert x)$ and $g(u,y)$ are not known in closed form, and, similar to the point-to-point rate--distortion function, the Wyner--Ziv rate--distortion function in Eq.~\eqref{eq:WZ} has a closed-form expression only in a few special cases.

The literature contains several constructive codes for the Wyner--Ziv problem, which assume synthetic setups and specific correlation settings. Zamir et al.~\cite{zamir_ITW, zamir_TIT} provide optimal constructive mechanisms for binary and Gaussian sources, respectively, using nested linear and lattice codes. The authors of DIstributed Source Coding Using Syndromes (DISCUS)~\cite{DISCUS} formulate the Wyner--Ziv setup as a dual quantizer--channel coding problem in the finite blocklength regime. In this setting, the source is first quantized based on its marginal density, and the quantization codebook space is divided into \emph{cosets}, determined by the virtual channel between the side information and the quantized source. Instead of transmitting the quantization index, the encoder sends the coset index, leading to further rate reduction. The decoder then disambiguates the coset index with the side information, recovers the quantization index, and estimates the source based on the deduced index and side information according to the distortion criterion. This systematic partitioning of the quantized source space with cosets resembles the random binning procedure seen in the proofs of the Slepian--Wolf and Wyner--Ziv theorems that consider the asymptotic blocklength regime~\cite{Slepian_Wolf, Wyner_Ziv}. The proposed method achieves performance close to the asymptotic Wyner--Ziv rate--distortion bound, for Gaussian sources and linear correlation structures.

\section{Learned Data Compression} ~\label{sec:NTC}

The general problem of optimal quantization in high-dimensional spaces, which is the case for many practical compression problems, is known to be intractable without further constraints~\cite{Gersho_Gray}. Consequently, many current practical compressors rely on a technique called \emph{transform coding}~\cite{Goyal}, which operates by linearly transforming the data vector into a continuous-valued representation, quantizing its elements independently, and subsequently encoding the induced discrete representation using a lossless entropy code~\cite{picture_coding}.

The conventional wisdom once considered designing codes with nonlinear transforms as computationally infeasible. However, this premise has changed with the resurgence of ANNs, which can approximate arbitrary functions when equipped with the right set of parameters~\cite{Leshno}. Leveraging these tools for function approximation in the context of source coding, the popular class of neural compression methods can now be framed through the perspective of \emph{nonlinear transform coding} (NTC)~\cite{BalleJournal}.

Classic learned compression literature typically focuses on improving point-to-point (with no side information) rate--distortion performances of popular practical information sources, such as Kodak image dataset~\cite{kodak_dataset}. To this end, NTC seeks to find an \emph{analysis} transform, $g_{a}$, that maps the space of source values to the \emph{latent} space, and a \emph{synthesis} transform, $g_{s}$, that maps from the latent space back into the source/reconstruction space. Typically, the latent variables, $\mathbf{u}$, are quantized with uniform scalar quantization (i.e., rounding to the nearest integers) that yields $\hat{\mathbf{u}} = \nint{\mathbf{u}} $. The quantized values are then entropy coded using likelihoods generated from an entropy model, $p_{\hat{\mathbf{u}}}(\hat{\mathbf{u}})$. The goal is to minimize the operational rate--distortion trade-off, in combination with stochastic optimization methods (e.g., stochastic gradient descent), which can be compactly expressed by the following objective function: 
\begin{align}
\mathbb{E}_{\mathbf{x}}[-\log_{2}p_{\hat{\mathbf{u}}}(\nint{g_{a}(\mathbf{x})})]
   & + \: \lambda \cdot \mathbb{E}_{\mathbf{x}}[\mathrm{d}\left( \mathbf{x}, g_{s}(\nint{g_{a}(\mathbf{x})}\right)], \label{eq:lagrangian}
\end{align}
where $\lambda > 0$ controls the trade-off, and $\mathrm{d}(\cdot, \cdot)$ denotes some distortion measure that quantifies the discrepancy of the reconstructed vector with respect to the input source. The transforms, $g_{a}$ and $g_{s}$, as well as the entropy model $p_{\hat{\boldsymbol{u}}}(\hat{\boldsymbol{u}})$ are all parametrized using neural networks. In the NTC case, unlike traditional transform coding that involves handcrafted transformations, the analysis and synthesis transforms are equipped with nonlinear transforms (e.g., variations of the rectified linear unit activation function). These transforms, along with the entropy model, can be optimized for any differentiable distortion measure $\mathrm{d}(\cdot, \cdot)$ in an end-to-end fashion by building onto the rate--distortion Lagrangian in Eq.~\eqref{eq:lagrangian}.

Note that the rounding operation in Eq.~\eqref{eq:lagrangian} is non-differentiable, which poses a challenge for optimizing the encoder. This means that optimizing the parameters of the analysis function, $g_{a}$, directly via \emph{backpropagation}, a crucial component of the learning process that neural methods rely on, is not possible. To address this, a common strategy is to substitute quantization with a differentiable approximation during training, typically through additive uniform noise, while retaining the quantization as is during inference time~\cite{Balle2017}. Recently, training lossy compressors without relying on any quantization approximations has gained popularity in the compression community~\cite{agustsson2020universally}, which draws inspiration from the literature on \emph{reverse channel coding}, also known as \emph{channel simulation}~\cite{Reverse_Shannon}. Instead of using additive uniform noise as a differentiable proxy during training, this line of work advocates for a differentiable channel that is also directly employed during inference time, in place of switching to hard quantization~\cite{agustsson2020universally}. Although the mismatch between training and inference phases is eliminated as such by foregoing quantization, it remains unclear how well-suited such methods are to practical compression scenarios as the computational complexity of the proposed methods grows exponentially with the number of bits to be transmitted~\cite{havasi2018minimal, agustsson2020universally, Flamich, communication_samples}.

\section{Distributed Compression with Real-Life Sources} \label{sec:dsc_real_life}

As explained in Section~\ref{sec:NTC}, conventional learned compression frameworks aim at enhancing point-to-point rate--distortion performance across various practical image datasets. While stereo image compression (jointly compressing the two images of a stereo pair by a single encoder) has already garnered attention from the compression community~\cite{stereo_MPEG, Liu_2019_ICCV}, DSC approaches, concerning independent encoding and joint decoding of correlated images, have remained vastly unexplored until recently. 

One of the early works that introduced a neural image compression framework fusing decoder-only side information is the approach proposed by Ayzik and Avidan~\cite{Ayzik_Avidan}. In this work, they first reconstruct the original image using a learned point-to-point compression framework, followed by additional refinement of the reconstructed image using side information image available at the decoder side. Hence, their method can be categorized as a ``post-processing" technique of the original image with the help of side information, as it does not rely on the side information image to compress the original one.

Exploiting the side information in a more efficient way has since been investigated in~\cite{NDIC_CAM, NDIC, NDSC}. Mital et al. employ cross-attention modules at the decoder side, in order to further align the features in~\cite{NDIC_CAM}, and explicitly model the common information between the input and side information in~\cite{NDIC}. The most recent work in~\cite{NDSC} uses a conditional Vector-Quantized Variational Autoencoder (VQ-VAE)~\cite{vq_vae}, in lieu of having a learned entropy model as in NTC~\cite{BalleJournal}, and establishes a conceptual link between the decoder-only side information setting and an optimization objective based on a variant of the evidence lower bound. The empirical results provided in~\cite{Ayzik_Avidan, NDIC_CAM, NDIC, NDSC} demonstrate that incorporating the decoder-only side information, in the form of a correlated image, into the compressor significantly improves the operational rate--distortion performance across various practical stereo image datasets, affirming that the neural compressors can successfully exploit the correlation between the images. The benefit of decoder-only side information has also been studied in the context of data-driven joint source channel coding (JSCC) for correlated image transmission~\cite{yilmaz2023distributed}. In this work, the authors demonstrated that the learned JSCC framework which exploits decoder-only side information outperforms both the learned point-to-point JSCC scheme~\cite{Bourtsoulatze_2019} and also the separation-based transmission scheme that makes use of side information by adopting a two-step source and channel coding approach. Recently, a learning-based multi-view image coding that involves independent encoding of correlated images followed by joint decoding was proposed in~\cite{zhang2023ldmic}.

Another line of research involving a learned DSC framework was explored in~\cite{li2023taskaware}, where the authors introduced a \emph{task-aware} distributed compression method focusing solely on downstream tasks, such as object detection, instead of reconstructing the original input source as in works that were discussed previously~\cite{Ayzik_Avidan, NDIC_CAM, NDIC, NDSC, yilmaz2023distributed, zhang2023ldmic}. The proposed objective function in this case is not based on a rate--distortion Lagrangian (as in Eq.~\eqref{eq:lagrangian}), but rather on minimizing the final composite task loss subjected to a fixed latent dimension that the authors term as the \emph{bandwidth}. Relying on \emph{principal component analysis} as a means of linear dimensionality reduction, which is coupled with a neural autoencoder aiming to extract task-relevant representations to be compressed, the approach learns to allocate respective bandwidths to each source based on the importance of the task at hand.

While learning-based methods offer a pathway to develop practical distributed compressors, which has been challenging with earlier approaches (Section~\ref{sec:theoretical_background}), relying on neural networks as a ``black box'' in real-world scenarios may have its drawbacks. As the explicit source distributions emerging in these aforementioned practical setups are unknown, the optimal rate--distortion bounds in such distributed compression scenarios are not formally characterized. Consequently, a comprehensive assessment of the actual performance of these proposed learned distributed compressors and a thorough understanding of their deviation from the theoretical optimum remain challenging. By examining exemplary abstract sources, for which the theoretical optimal performance is known (e.g., quadratic-Gaussian model in the Wyner--Ziv case), it becomes more feasible to gauge the effectiveness of neural distributed compressors and to understand whether they actually recover the optimum, or near optimal, solutions. Considering such abstract settings allows us to devise well-informed architectural design choices, which can also potentially improve compression efficiency for high-dimensional sources such as images.

\section{Learned Wyner--Ziv Coding with Abstract Sources} \label{sec:learned_wz}

Several works have investigated how learned compressors perform on abstract sources to gain a better understanding of the conditions that lead to optimal performance. The study in~\cite{BalleJournal} validated that the NTC-based models can achieve the same compression performance as the optimal (point-to-point) entropy-constrained scalar quantizer for the Laplace distribution, whose performance was established by Sullivan~\cite{Sullivan}. Similarly, Wagner and Ballé~\cite{wagner2020neural} showed that the NTC-based schemes can also optimally compress sources that exhibit a low-dimensional manifold structure in a high-dimensional ambient space, such as a specific random process named the \emph{sawbridge}. Notably, the recent work in~\cite{bhadane2022neural} demonstrated that the same class of learned compressors are, surprisingly, unable to optimally compress certain complex sources exhibiting a circular topology, such as a particular random process named the \emph{ramp}. The authors hypothesized that this is due to the NTC-based schemes having difficulty learning sufficiently steep functions, a phenomenon known as \emph{spectral bias}~\cite{rahaman2019spectral} in learning literature. These analyses offer valuable insights into the fundamental learning capabilities of ANN-based compressors, and have the potential to exert a significant impact on image compression. However, all  of these works only considered point-to-point information sources as inputs, without accounting for neither any presence  of side information nor any other distributed compression setup in their formulation.

As one of the initial studies on a learned approach for the Wyner--Ziv problem for abstract sources, the authors of~\cite{data_driven_wz} considered a data-driven iterative optimization scheme for Gauss--Markov source and side information sequences with arbitrary memory. Building their model structure onto the Wyner--Ziv coding architectures proposed in~\cite{Tuncel1, Tuncel2}, which involve one-dimensional scalar quantization and scalar binning of quantization indices, this work showcased that a lower-complexity data-driven method can match the performance of a quantizer optimized using an exhaustive search.

Inspired by lossy compression techniques rooted in reverse channel coding, the recent work in~\cite{phan2024importance} established a novel one-shot coding scheme for the Wyner--Ziv problem, where the authors introduced an \emph{importance sampling} based compression method that is a variant of random coding~\cite{communication_samples}. As such, this approach suggests entirely omitting the discretization/quantization step, in contrast to the NTC-based compressors (see Section~\ref{sec:NTC}), and instead assumes the existence of a shared source of randomness available to both the encoder and the decoder, similar to the recently introduced Poisson functional representation lemma~\cite{Li_2021}.

\begin{figure}
    \centering
 \includegraphics[scale=0.22]{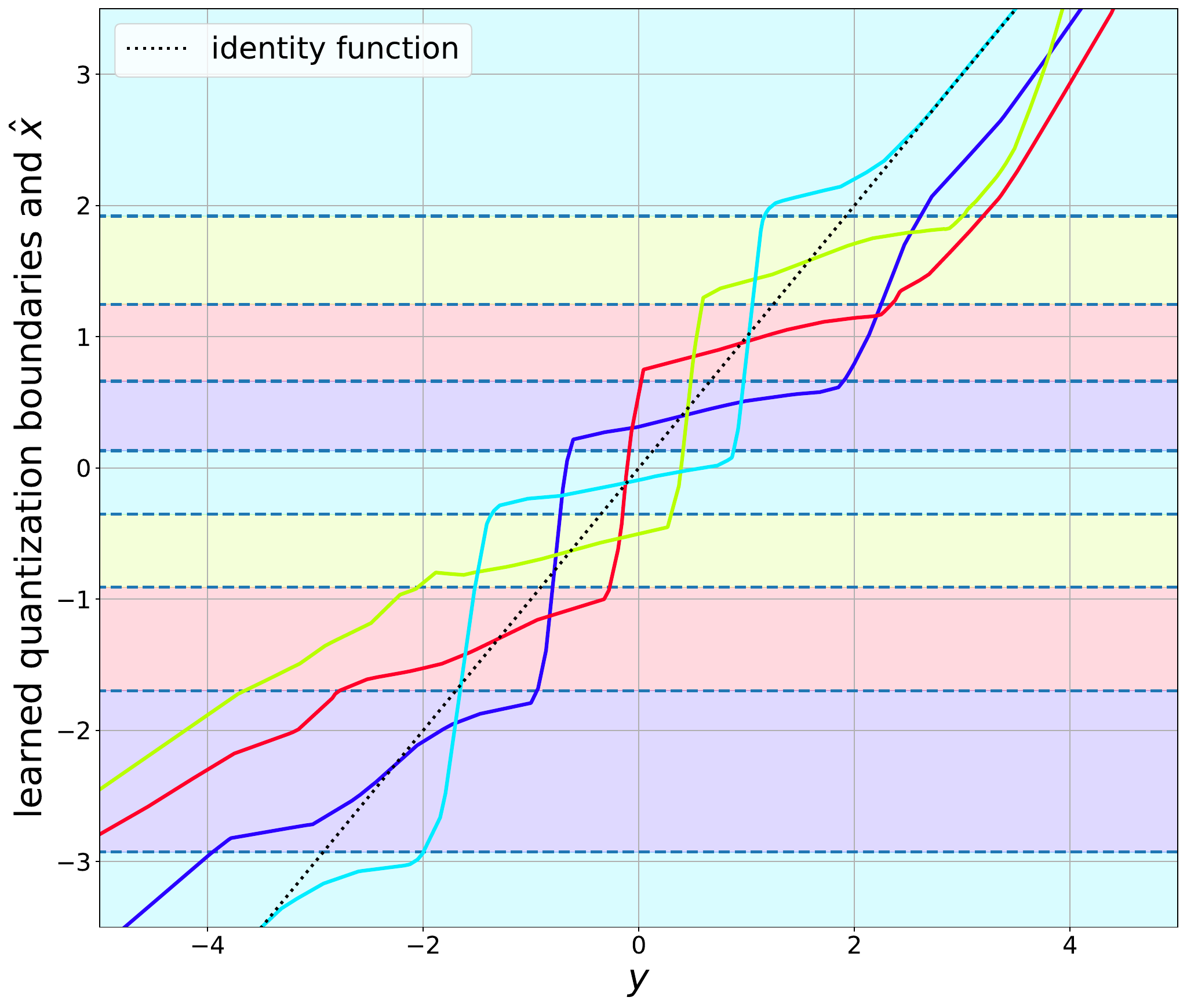}
\caption{Figure taken from~\cite{ozyilkan2023learned}. Visualization (best viewed in color) of the learned encoder and decoder of the operational scheme proposed in~\cite{ozyilkan2023learned}, for the quadratic-Gaussian Wyner--Ziv setup. The dashed horizontal lines are quantization boundaries, and the colors between boundaries represent unique values of bin indices. The decoding function is depicted as separate plots for each value of bin index, using the same color assignment. Color coding reveals that the model learns discontiguous quantization bins.}
    \label{fig:isit_23}
\end{figure}

While the NTC-based compressors seem to be a good candidate for the Wyner--Ziv problem, our analysis in~\cite{ozyilkan2023neural} revealed that surprisingly, they fail to exploit the side information most efficiently. We argued that this popular class of neural compressors are unable to recover any flexible and efficient binning schemes even considering a simple test case, where side information is the sign function of input realization. As an alternative to the NTC-based compressors, our work in~\cite{ozyilkan2023neural, ozyilkan2023neuralworkshop, ozyilkan2023learned} proposed a more generic learning-based algorithm, which represents the first unstructured entropy-constrained vector quantizer that makes use of side information. We demonstrated that such a learned compressor can re-discover some principles of the optimum theoretical solution for the Wyner--Ziv problem, such as optimal combination of the quantization index and side information at the decoder, for various exemplary sources. In particular, we illustrated that these learned distributed compressors are capable of exhibiting different types of interpretable binning mechanisms, such as periodic-like mappings (see Fig.~\ref{fig:isit_23}) for the quadratic-Gaussian case and symmetric mappings for the sign function as side information, akin to the achievability part of the Wyner--Ziv theorem (see Section~\ref{sec:WZ}). This provides empirical evidence that the ANN-based methods can learn constructive solutions very similar to some of the handcrafted compressors proposed for the Wyner--Ziv problem, such as DISCUS~\cite{DISCUS}, without requiring a priori knowledge of source statistics. We also justify how these proposed models are interpretable as operational schemes by picking a suitable entropy coding technique for each. These learned distributed compressors, coupled with a high-order entropy coding scheme (either classic or Slepian--Wolf) yield a better performance compared to the point-to-point rate--distortion functions, and in some cases achieve the optimal information-theoretic entropy--distortion bounds with side information~\cite{ozyilkan2023neural}. We note that our proposed entropy-constrained compression frameworks are more amenable to practical use, in comparison to the scheme proposed in~\cite{phan2024importance} that is based on communicating samples from continuous distributions. However, in their current form, the operational schemes in~\cite{ozyilkan2023neural, ozyilkan2023neuralworkshop, ozyilkan2023learned} inherit the limitations of vector quantization, that is naïvely increasing the codebook size renders these methods computationally infeasible.

\section{Discussion} \label{sec:conclusion}

In this article, we have reviewed the literature on learned distributed compressors, with a specific focus on the schemes that assessed their compression performance on abstract sources. Recent findings demonstrate that ANN-based compressors, even without the explicit knowledge of source distributions, have emerged as competitive candidates for addressing open problems in distributed compression and offer promising constructive solutions to long-standing challenges in the field, such as adaptability to complex correlations.

Possible avenues for future work include analyzing the robustness of these neural schemes across various correlation structures and also extending them to fully distributed compression settings. Exploring this research direction has the potential to greatly enhance compression efficiency for transmitting practical signals in communication networks, such as multi-view image and video. The development of learned distributed compressors can also play a role in the practical design of fundamental building blocks of cooperative communications, such as the relay channel~\cite{relaycapacity}.

\section*{Acknowledgments}
We would like to thank Johannes Ballé for helpful discussions and valuable comments on this manuscript. 

\bibliographystyle{IEEEtran}

\bibliography{ref.bib}

\begin{thebibliography}{10}
\providecommand{\url}[1]{#1}
\csname url@samestyle\endcsname
\providecommand{\newblock}{\relax}
\providecommand{\bibinfo}[2]{#2}
\providecommand{\BIBentrySTDinterwordspacing}{\spaceskip=0pt\relax}
\providecommand{\BIBentryALTinterwordstretchfactor}{4}
\providecommand{\BIBentryALTinterwordspacing}{\spaceskip=\fontdimen2\font plus
\BIBentryALTinterwordstretchfactor\fontdimen3\font minus \fontdimen4\font\relax}
\providecommand{\BIBforeignlanguage}[2]{{%
\expandafter\ifx\csname l@#1\endcsname\relax
\typeout{** WARNING: IEEEtran.bst: No hyphenation pattern has been}%
\typeout{** loaded for the language `#1'. Using the pattern for}%
\typeout{** the default language instead.}%
\else
\language=\csname l@#1\endcsname
\fi
#2}}
\providecommand{\BIBdecl}{\relax}
\BIBdecl

\bibitem{Shannon1}
C.~E. Shannon, ``A mathematical theory of communication,'' \emph{The Bell System Technical Journal}, vol.~27, no.~3, pp. 379--423, 1948.

\bibitem{Shannon2}
------, \emph{Coding Theorems for a Discrete Source With a Fidelity CriterionInstitute of Radio Engineers, International Convention Record, vol. 7, 1959.}, 1993, pp. 325--350.

\bibitem{elements_of_information_theory}
T.~M. Cover and J.~A. Thomas, \emph{Elements of Information Theory (Wiley Series in Telecommunications and Signal Processing)}.\hskip 1em plus 0.5em minus 0.4em\relax USA: Wiley-Interscience, 2006.

\bibitem{JPEG}
G.~Wallace, ``The {JPEG} still picture compression standard,'' \emph{IEEE Transactions on Consumer Electronics}, vol.~38, no.~1, pp. xviii--xxxiv, 1992.

\bibitem{Balle2017}
J.~Ball\'{e}, V.~Laparra, and E.~P. Simoncelli, ``End-to-end optimized image compression,'' in \emph{International Conference on Learning Representations}, 2017.

\bibitem{Balle2018}
J.~Ballé, D.~Minnen, S.~Singh, S.~J. Hwang, and N.~Johnston, ``Variational image compression with a scale hyperprior,'' in \emph{International Conference on Learning Representations}, 2018.

\bibitem{BalleJournal}
J.~Ballé, P.~A. Chou, D.~Minnen, S.~Singh, N.~Johnston, E.~Agustsson, S.~J. Hwang, and G.~Toderici, ``Nonlinear transform coding,'' \emph{IEEE Journal of Selected Topics in Signal Processing}, vol.~15, no.~2, pp. 339--353, 2021.

\bibitem{Minnen2018}
D.~C. Minnen, J.~Ball{\'e}, and G.~Toderici, ``Joint autoregressive and hierarchical priors for learned image compression,'' in \emph{Neural Information Processing Systems}, 2018.

\bibitem{BPG}
F.~Bellard, ``{BPG} image format,'' \url{https://bellard.org/bpg/}, 2014.

\bibitem{Leshno}
\BIBentryALTinterwordspacing
M.~Leshno, V.~Y. Lin, A.~Pinkus, and S.~Schocken, ``Multilayer feedforward networks with a nonpolynomial activation function can approximate any function,'' \emph{Neural Networks}, vol.~6, no.~6, pp. 861--867, Jan. 1993. [Online]. Available: \url{https://doi.org/10.1016/s0893-6080(05)80131-5}
\BIBentrySTDinterwordspacing

\bibitem{Wyner_Ziv}
A.~Wyner and J.~Ziv, ``The rate--distortion function for source coding with side information at the decoder,'' \emph{IEEE Transactions on Information Theory}, vol.~22, no.~1, pp. 1 -- 10, 1976.

\bibitem{Slepian_Wolf}
D.~Slepian and J.~Wolf, ``Noiseless coding of correlated information sources,'' \emph{IEEE Transactions on Information Theory}, vol.~19, no.~4, pp. 471 -- 480, 1973.

\bibitem{network_info_theo}
A.~E. Gamal and Y.-H. Kim, \emph{Network Information Theory}.\hskip 1em plus 0.5em minus 0.4em\relax USA: Cambridge University Press, 2012.

\bibitem{Cover1}
T.~Cover, ``A proof of the data compression theorem of {S}lepian and {W}olf for ergodic sources (corresp.),'' \emph{IEEE Transactions on Information Theory}, vol.~21, no.~2, pp. 226--228, 1975.

\bibitem{multiterminal}
T.~Berger, \emph{Multiterminal Source Coding. In G. Longo (Ed.), The Information Theory Approach to Communications}.\hskip 1em plus 0.5em minus 0.4em\relax USA: New York, NY, USA: Springer-Verlag, 1977.

\bibitem{Aaron_quadratic}
A.~B. Wagner, S.~Tavildar, and P.~Viswanath, ``Rate region of the quadratic gaussian two-encoder source-coding problem,'' in \emph{2006 IEEE International Symposium on Information Theory}, 2006, pp. 1404--1408.

\bibitem{zamir_ITW}
R.~Zamir and S.~Shamai, ``Nested linear/lattice codes for {W}yner-{Z}iv encoding,'' in \emph{1998 Information Theory Workshop (Cat. No.98EX131)}, 1998, pp. 92--93.

\bibitem{zamir_TIT}
R.~Zamir, S.~Shamai, and U.~Erez, ``Nested linear/lattice codes for structured multiterminal binning,'' \emph{IEEE Transactions on Information Theory}, vol.~48, no.~6, pp. 1250--1276, 2002.

\bibitem{DISCUS}
S.~Pradhan and K.~Ramchandran, ``Distributed source coding using syndromes ({DISCUS}): {D}esign and construction,'' \emph{IEEE Transactions on Information Theory}, vol.~49, no.~3, pp. 626--643, 2003.

\bibitem{Gersho_Gray}
A.~Gersho and R.~M. Gray, \emph{Vector quantization and signal compression}.\hskip 1em plus 0.5em minus 0.4em\relax USA: Kluwer Academic Publishers, 1991.

\bibitem{Goyal}
V.~Goyal, ``Theoretical foundations of transform coding,'' \emph{IEEE Signal Processing Magazine}, vol.~18, no.~5, pp. 9--21, 2001.

\bibitem{picture_coding}
A.~Netravali and J.~Limb, ``Picture coding: A review,'' \emph{Proceedings of the IEEE}, vol.~68, no.~3, pp. 366--406, 1980.

\bibitem{kodak_dataset}
E.~Kodak, ``Kodak lossless true color image suite ({P}hoto{CD} {PCD}0992),'' \url{http://r0k.us/graphics/kodak}.

\bibitem{agustsson2020universally}
E.~Agustsson and L.~Theis, ``Universally quantized neural compression,'' in \emph{Proceedings of the 34th International Conference on Neural Information Processing Systems}, 2020.

\bibitem{Reverse_Shannon}
C.~Bennett, P.~Shor, J.~Smolin, and A.~Thapliyal, ``Entanglement-assisted capacity of a quantum channel and the reverse shannon theorem,'' \emph{IEEE Transactions on Information Theory}, vol.~48, no.~10, pp. 2637--2655, 2002.

\bibitem{havasi2018minimal}
M.~Havasi, R.~Peharz, and J.~M. Hernández-Lobato, ``Minimal random code learning: Getting bits back from compressed model parameters,'' in \emph{International Conference on Learning Representations}, 2019.

\bibitem{Flamich}
G.~Flamich, M.~Havasi, and J.~M. Hern\'{a}ndez-Lobato, ``Compressing images by encoding their latent representations with relative entropy coding,'' in \emph{Proceedings of the 34th International Conference on Neural Information Processing Systems}, 2020.

\bibitem{communication_samples}
L.~Theis and N.~Y. Ahmed, ``Algorithms for the communication of samples,'' in \emph{Proceedings of the 39th International Conference on Machine Learning}, ser. Proceedings of Machine Learning Research, vol. 162.\hskip 1em plus 0.5em minus 0.4em\relax PMLR, 17--23 Jul 2022, pp. 21\,308--21\,328.

\bibitem{stereo_MPEG}
P.~Merkle, K.~Muller, A.~Smolic, and T.~Wiegand, ``Efficient compression of multi-view video exploiting inter-view dependencies based on {H.264/MPEG4-AVC},'' in \emph{2006 IEEE International Conference on Multimedia and Expo}, 2006, pp. 1717--1720.

\bibitem{Liu_2019_ICCV}
J.~Liu, S.~Wang, and R.~Urtasun, ``{DSIC}: Deep stereo image compression,'' in \emph{Proceedings of the IEEE/CVF International Conference on Computer Vision}, October 2019.

\bibitem{Ayzik_Avidan}
S.~Ayzik and S.~Avidan, ``Deep image compression using decoder side information,'' in \emph{Computer Vision – ECCV 2020: 16th European Conference, Glasgow, UK, August 23–28, 2020, Proceedings, Part XVII}.\hskip 1em plus 0.5em minus 0.4em\relax Berlin, Heidelberg: Springer-Verlag, 2020, p. 699–714.

\bibitem{NDIC_CAM}
N.~Mital, E.~Ozyilkan, A.~Garjani, and D.~Gunduz, ``Neural distributed image compression with cross-attention feature alignment,'' in \emph{2023 IEEE/CVF Winter Conference on Applications of Computer Vision}, Jan 2023, pp. 2497--2506.

\bibitem{NDIC}
N.~Mital, E.~Özyılkan, A.~Garjani, and D.~Gündüz, ``Neural distributed image compression using common information,'' in \emph{2022 Data Compression Conference}, 2022, pp. 182--191.

\bibitem{NDSC}
\BIBentryALTinterwordspacing
J.~Whang, A.~Nagle, A.~Acharya, H.~Kim, and A.~G. Dimakis, ``Neural distributed source coding,'' 2023. [Online]. Available: \url{https://arxiv.org/abs/2106.02797}
\BIBentrySTDinterwordspacing

\bibitem{vq_vae}
A.~van~den Oord, O.~Vinyals, and K.~Kavukcuoglu, ``Neural discrete representation learning,'' in \emph{Proceedings of the 31st International Conference on Neural Information Processing Systems}, 2017, p. 6309–6318.

\bibitem{yilmaz2023distributed}
\BIBentryALTinterwordspacing
S.~F. Yilmaz, E.~Ozyilkan, D.~Gunduz, and E.~Erkip, ``Distributed deep joint source-channel coding with decoder-only side information,'' to appear in \emph{2024 IEEE International Conference on Machine Learning for Communication and Networking}, 2023. [Online]. Available: \url{https://arxiv.org/abs/2310.04311}
\BIBentrySTDinterwordspacing

\bibitem{Bourtsoulatze_2019}
E.~Bourtsoulatze, D.~Burth~Kurka, and D.~Gunduz, ``Deep joint source-channel coding for wireless image transmission,'' \emph{IEEE Transactions on Cognitive Communications and Networking}, vol.~5, no.~3, p. 567–579, Sep. 2019.

\bibitem{zhang2023ldmic}
X.~Zhang, J.~Shao, and J.~Zhang, ``{LDMIC}: Learning-based distributed multi-view image coding,'' in \emph{The Eleventh International Conference on Learning Representations}, 2023.

\bibitem{li2023taskaware}
P.~Li, S.~K. Ankireddy, R.~Zhao, H.~N. Mahjoub, E.~M. Pari, U.~Topcu, S.~P. Chinchali, and H.~Kim, ``Task-aware distributed source coding under dynamic bandwidth,'' in \emph{Thirty-seventh Conference on Neural Information Processing Systems}, 2023.

\bibitem{Sullivan}
G.~Sullivan, ``Efficient scalar quantization of exponential and {L}aplacian random variables,'' \emph{IEEE Transactions on Information Theory}, vol.~42, no.~5, pp. 1365--1374, 1996.

\bibitem{wagner2020neural}
A.~B. Wagner and J.~Ballé, ``Neural networks optimally compress the sawbridge,'' in \emph{2021 Data Compression Conference}, 2021, pp. 143--152.

\bibitem{bhadane2022neural}
S.~Bhadane, A.~B. Wagner, and J.~Ballé, ``Do neural networks compress manifolds optimally?'' in \emph{2022 IEEE Information Theory Workshop}, 2022, pp. 582--587.

\bibitem{rahaman2019spectral}
N.~Rahaman, A.~Baratin, D.~Arpit, F.~Draxler, M.~Lin, F.~Hamprecht, Y.~Bengio, and A.~Courville, ``On the spectral bias of neural networks,'' in \emph{Proceedings of the 36th International Conference on Machine Learning}, vol.~97, 2019, pp. 5301--5310.

\bibitem{data_driven_wz}
E.~Domanovitz, D.~Severo, A.~Khisti, and W.~Yu, ``Data-driven optimization for zero-delay lossy source coding with side information,'' in \emph{ICASSP 2022 - 2022 IEEE International Conference on Acoustics, Speech and Signal Processing}, 2022, pp. 5203--5207.

\bibitem{Tuncel1}
E.~Tuncel, ``Predictive coding of correlated sources,'' in \emph{Information Theory Workshop}, 2004, pp. 111--116.

\bibitem{Tuncel2}
X.~Chen and E.~Tuncel, ``Low-delay prediction- and transform-based {W}yner–{Z}iv coding,'' \emph{IEEE Transactions on Signal Processing}, vol.~59, no.~2, pp. 653--666, 2011.

\bibitem{phan2024importance}
\BIBentryALTinterwordspacing
B.~Phan, A.~Khisti, and C.~Louizos, ``Importance matching lemma for lossy compression with side information,'' 2024. [Online]. Available: \url{https://arxiv.org/abs/2401.02609}
\BIBentrySTDinterwordspacing

\bibitem{Li_2021}
C.~T. Li and V.~Anantharam, ``A unified framework for one-shot achievability via the {P}oisson matching lemma,'' \emph{{IEEE} Transactions on Information Theory}, vol.~67, no.~5, pp. 2624--2651, May 2021.

\bibitem{ozyilkan2023learned}
E.~Özyılkan, J.~Ballé, and E.~Erkip, ``Learned {W}yner–{Z}iv compressors recover binning,'' in \emph{2023 IEEE International Symposium on Information Theory (ISIT)}, 2023, pp. 701--706.

\bibitem{ozyilkan2023neural}
\BIBentryALTinterwordspacing
E.~Ozyilkan, J.~Ballé, and E.~Erkip, ``Neural distributed compressor discovers binning,'' 2023. [Online]. Available: \url{https://arxiv.org/abs/2310.16961}
\BIBentrySTDinterwordspacing

\bibitem{ozyilkan2023neuralworkshop}
E.~Ozyilkan, J.~Ball{\'e}, and E.~Erkip, ``Neural distributed compressor does binning,'' in \emph{ICML 2023 Workshop Neural Compression: From Information Theory to Applications}, 2023.

\bibitem{relaycapacity}
T.~Cover and A.~Gamal, ``Capacity theorems for the relay channel,'' \emph{IEEE Transactions on Information Theory}, vol.~25, no.~5, pp. 572--584, 1979.

\end{thebibliography}

\newpage

\end{document}